\documentclass{JHEP}
\usepackage{amsfonts,amsmath}
\newcommand{\ii}{\mathrm{i}}
\newcommand{\na}{\nabla}
\newcommand{\dd}{\mathrm{d}}
\newcommand{\pd}{\partial}
\newcommand{\hh}{\mathcal{H}}

\newcommand{\e}{\mathrm{e}}
\newcommand{\ket}[1]{\left|#1\right\rangle}
\newcommand{\bra}[1]{\left\langle #1\right|}

\newcommand{\tr}{\mathop{\mathrm{tr}}}

\title{Brane Vacuum as Chain of Rotators}%

\author{E.~E.~Donets$^{a,b}$, A.~P.~Isaev$^a$, C.~Sochichiu$^{a,c}$,
and M.~Tsulaia$^{a,d}$\\
$^a$Bogoliubov Laboratory of Theoretical Physics,\\
Joint Institute for Nuclear Research\\ 141980 Dubna, Moscow Reg.\\
RUSSIA\\
$^b$Laboratory of High Energies,\\
Joint Institute for Nuclear Research\\ 141980 Dubna, Moscow Reg.\\
RUSSIA\\
$^c$Institutul de Fizic\u a Aplicat\u a A\c S,\\
 str. Academiei, nr. 5, Chi\c sin\u au,\\
MD2028 MOLDOVA\\
$^d$Andronikashvili Institute of Physics\\ 380077 Tbilisi, GEORGIA \\
\email{edonets@sunhe.jinr.ru}\\
\email{isaevap@thsun1.jinr.ru}\\
\email{sochichi@thsun1.jinr.ru}\\
\email{tsulaia@thsun1.jinr.ru}}

\preprint{\hepth{0011090}}

\abstract{
 We analyse the noncommutative U(1) sigma model, which arises from the
 vacuum dynamics of the noncommutative charged tachyonic field.
 The sector of ``spherically symmetric'' excitations of the model is equivalent
 to a chain of rotators. Classical solutions for this model are found,
 which are static and ``spherically symmetric'' in noncommutative spatial
 dimensions. The limit of small noncommutativity
 reveals the presence of Polyakov vortices in the model.
 A generalisation of the model to $q$-deformed space, which
 may serve as a regularisation of the non-deformed model is also considered.
}
\begin{document}
\section{Introduction}

The noncommutative geometry \cite{Connes:2000by}, proved to be an utile tool in the
brane dynamics of string theory \cite{Seiberg:1999vs}.

Thus, the low energy effective action for a $p$-brane in the presence of nonzero
constant $B_{\mu\nu}$ field along the brane is given by the dimensional reduction
of the 10-dimensional noncommutative Yang--Mills model to the brane world volume
\cite{Connes:1998cr}.

In the limit of large field $B_{\mu\nu}$, i.e. when by some matrix norm,
\begin{equation}\label{large_B}
  \alpha'\|B_{\mu\nu}\|\gg
  \|g_{\mu\nu}\|,
\end{equation}
 the noncommutativity parameter $\theta_{\mu\nu}$ is given by
the inverse matrix to $B$ \cite{Seiberg:1999vs},
\begin{equation}
   \theta^{\mu\nu}=(B^{-1})^{\mu\nu},
\end{equation}
while the open string metric along the brane is given by,
\begin{equation}\label{open_str_g}
  G_{\mu\nu}=-(2\pi\alpha')^2(Bg^{-1}B)_{\mu\nu},
\end{equation}
and the Yang--Mills coupling,
\begin{equation}\label{YM_coupl}
  g^2_{YM}=\frac{(2\pi)^{p-2}G_s}{(\alpha')^{\frac{3-p}{2}}},
\end{equation}
where, $G_s$ is the open string coupling. In this limit the stringy corrections
decouple and the model is described by the effective noncommutative theory
\cite{Seiberg:1999vs,Witten:2000nz}.

From the other hand, in the case of a brane-antibrane pair or of a non-BPS
non-stable brane one finds in the spectrum of the effective theory tachyonic modes
as well. In the above limit (\ref{large_B}), the tachyonic mode is described by
noncommutative scalar field(s) $T$ with potential $V(T)$ (tachyonic potential)
\cite{Sen:1999md,Sen:1999xm,Sen:2000vx}.

The action for brane system containing a tachyonic mode in this case looks as one
for the noncommutative scalar field interacting with U(1) gauge field
\cite{Harvey:2000jt,Dasgupta:2000ft},
\begin{equation}\label{tach_action}
  S_{\mathrm{Brane}}=\int \dd^{p+1}x
  \left(-\frac{1}{4g^2_{YM}}\eta^{\mu\alpha}\eta^{\nu\beta}
  F_{\mu\nu}*F_{\alpha\beta}+ \eta^{\mu \nu}
  \frac{1}{2}f(T)*\na_\mu T*\na_\nu T-
  V(*T)\right),
\end{equation}
where $\eta _{\mu \alpha}=\mathrm{diag}(1,-1,...,-1)$ and we put the brane tension
equal to unity. In the above equation all products are understood as star products,
defined by
\begin{equation}\label{star}
  A *B(x)=
  \e^{\frac{\ii}{2}\theta^{\mu\nu}\pd_\mu\pd_\nu'}A(x) B(x')
  \big|_{x'=x}.
\end{equation}
The star product can be viewed as one of possible representation of the algebra of
operators acting on some Hilbert space $\hh$, integral given by the trace.

The tachyonic potential is known to have a local maximum at the origin and global
minimum for the some constant, i.e. $c$-number value $T_*$ of the field. In the
case of superstrings the potential is even with respect to the inversion $T\to -T$,
and there is also minimum at $-T_*$.

In the case of zero gauge fields one can find a solution to the equations of
motion of the tachyon $T$ in the limit of the large noncommutativity
parameter\footnote{This limit, however, should be consistent with, eq.
(\ref{large_B}).} $\theta$ when the kinetic term is negligible in comparison with
potential one in the equations of motion \cite{Gopakumar:2000zd}. Such a solution
is given by $T_0=T_*P(x)$, where $P(x)$ is the noncommutative function
corresponding to some projector $P*P=P$.

In the presence of the nontrivial gauge field one can find solution of the similar
type as above \cite{Sochichiu:2000rm,Gopakumar:2000rw}. In this case the solution
implies such gauge field configurations for which the tachyonic soliton is
covariantly constant,
\begin{equation}\label{cov_const}
  \na_\mu T_0=0,
\end{equation}
and requirement for noncommutativity parameter $\theta$ to be large is no more
needed.

The meaning of these solutions consists in the description of the smaller branes
living on the original $p$-brane \cite{Aganagic:2000xx}.

To obtain the solutions one uses only the minima of the potential and the fact that
the tachyonic field is covariantly constant, but not the details of the kinetic
term and interaction potential. If, however, one is interested in the spectrum of
fluctuations around such a vacuum solution these details may become important.

In this paper we introduce the model of the noncommutative U(1)-field having some
relation to the tachyonic vacuum. We find and analyse solutions which neither
appear in the commutative case nor fall in the class of solutions of the described
above projector type.

The plan of the paper is the following. In the next section we consider a model of
noncommutative charged Higgs-like field whose vacuum dynamics yields a
noncommutative sigma model ($U$-field). We give some kind of string motivation for
the model under consideration, however, the model itself presents a separate
interest. In the third section we analyse equations of motion, local extrema, and
geometrical meaning of the $U$-field. In the fourth section we extend these
results to the model living on $q$-deformed space, which can be treated as a
regularisation of the model above. In the fifth section we show, that in the
commutative limit one can treat $U$-field as a free scalar field surrounded with
the gas of Polyakov vortices. Finally, we discuss our results.

\section{Charged Tachyon}
Consider the noncommutative Higgs-like model of charged scalar fields (charged
noncommutative tachyon) in the trivial noncommutative gauge field background. It is
given by the action,
\begin{equation}\label{scalar}
  S=\int \dd^{p+1}x
  \left(\frac{1}{2} \eta^{\mu \nu}\pd_\mu \phi*\pd_\nu\phi^\dag
  -V(\phi*\phi^\dag)\right),
\end{equation}
where $V(\cdot)$ is a potential with a nontrivial v.e.v.: $|\phi|^2=$ some
constant.

The field $\phi$ transforms in the bi-fundamental representation of the U(1) gauge
group. Such fields were considered earlier in \cite{Harvey:2000jb}.

In general, (bi-)fundamental modes arise in the spectrum of the nonstable branes
\cite{Harvey:2000jt,Gopakumar:2000rw}. These modes correspond to strings which
have only one end on the brane. Since in the compactified theory a string cannot
end elsewhere it was conjectured by \cite{Sen:1999md,Sen:2000}, that such  modes
are absent. The ``confinement'' mechanism for these modes is as follows. Close to
the tachyonic vacuum the kinetic part of the noncommutative gauge field vanishes
and the gauge potential starts to play the role of the Lagrange multiplier for the
charged currents forcing them to vanish.

In the case of non-compact Minkowski space there is also the possibility to have
strings with one end at the infinity. Although the energy cost of such strings is
large their presence may change the situation. Indeed, if there is some nontrivial
v.e.v. for the charged field, which correspond to a tachyonic mode of the
semi-infinite strings, the above described mechanism does not freeze all the
degrees of freedom, since there remain ones connected with gauge motion along the
vacuum valley.

A point in the (true) vacuum of the field $\phi$ can be parameterised by an
element of the noncommutative U(1),
\begin{align}\label{U}
  &\phi\to U*\phi, \\
  &\phi^\dag \to \phi^\dag* U^{-1},
\end{align}
where $U$ and $U^{-1}$ satisfy $U*U^{-1}=U^{-1}*U=1$. $U$ act transitively along
the vacuum valley.

To get the action describing the dynamics along the valley of the potential $V$ in
terms of field $U$, let us take a constant $\phi$ realising the minimum of the
potential. After that, perform the transformation (\ref{U}) and declare the field
$U$ dynamical. The action for the Goldstone $U$-field looks as follows,
\begin{equation}\label{action_U}
  S
  =\frac{1}{\lambda^2}\int\dd^{p+1} x\, \eta^{\mu \nu}
   \pd_\mu U*\pd_\nu U^{-1}
  =-\frac{1}{\lambda^2}\int\dd^{p+1} x\, \eta^{\mu \nu}
   \pd_\mu U*U^{-1}*\pd_\nu U*U^{-1},
\end{equation}
where the coupling $\lambda$ is given by
 $1/\lambda^2=\phi*\phi^\dag=$constant.

The model (\ref{action_U}) describes the tachyon fluctuations along the degenerate
vacuum. In the limit of large $\theta$ the remaining fluctuations decouple and the
action (\ref{action_U}) describes all the tachyonic degrees of freedom. In what
follows we do not assume large or small value of the noncommutative parameter,
except the Section 5 where we consider the limit of \emph{small} $\theta$.

Let us note that a model of U(1)-field was considered in \cite{Gorsky:2000wy}, in
the limit of strong noncommutativity when one can neglect the kinetic term.

\section{The noncommutative $U$-field}

Consider the case $p=2$. We assume that two spatial coordinates
are noncommutative,
\begin{equation}\label{noncom}
  [x^1,x^2]=\ii \theta,
\end{equation}
while time coordinate is commutative. In this paper we assume Minkowski signature
and the commutative time. For definiteness, consider $\theta
>0$, the case of negative $\theta$ can be reduced to one under consideration by
interchange, of $x^1$ and $x^2$.

Equations of motion corresponding to the action (\ref{action_U}) look as follows,
\begin{multline}\label{EqM}
  -\pd^2 U^{-1}+U^{-1}*\pd^2 U* U^{-1}\equiv \\
  -\ddot{U}{}^{-1}+U^{-1}*\ddot{U}*U^{-1}+\pd_i^2 U^{-1}-U^{-1}*\pd_i^2 U* U^{-1}=0,
\end{multline}
where $i=1,2$.

Consider this equation in the operator form, and let us search for classical
solutions. For this it is useful to pass to ``complex coordinates'' given by
oscillator representation, and work in terms of operators,
\begin{align}\label{aabar}
  & a=\frac{1}{\sqrt{2\theta}}(x^1+\ii x^2), \qquad
  \bar{a}=\frac{1}{\sqrt{2\theta}}(x^1-\ii x^2), \qquad N=\bar{a}a, \\
  & [a,\bar{a}]=1, \qquad Na=a(N-1),\qquad N\bar{a}=\bar{a}(N+1).
\end{align}

Assume the ``spherically symmetric'' ansatz $U(N,t)\equiv U_N(t)=\e^{\ii u_N(t)}$
(see e.g. \cite{Gopakumar:2000zd}), then the equation for the ansatz translates to,
\begin{equation}\label{ansatz}
    \ddot{u}_N = \frac{2}{\theta} \left[(N+1) \sin (u_{N+1} - u_N)  -
    N \sin (u_N  - u_{N-1} )  \right].
\end{equation}

The equations of motion can be integrated back to an action. The action one gets
coincides with simple reduction of the original action (\ref{action_U}) to the
ansatz,
\begin{equation}\label{action_ansatz}
  S = \frac{2\pi\theta}{\lambda^2}  \int_{-\infty}^{+\infty} \dd t  \tr
  \left\{(\dot{u}_N )^2  - \frac{4}{\theta} N  [  1 - \cos (u_N  -
  u_{N-1} ) ]\right\},
\end{equation}
It describes the ``spherically symmetric'' excitations in the model of $U$-field.
Further we will consider only the degrees of freedom of this type. The action
(\ref{action_ansatz}) and equations of motion (\ref{ansatz}) are given in the
operator form and we use the equivalence between noncommutative integrals and
traces,
\[
\int \dd^2x (\cdot) \equiv  (2\pi \theta)\tr (\cdot).
\]

Let us note that this generalises to the case of $p = 2k$ noncommutative
coordinates, where one has the action
\begin{equation}\label{action2k}
  S = \frac{(2\pi\theta)^k}{\lambda^2}  \int_{-\infty}^{+\infty} \dd t  \tr
  \left\{(\dot{u}_{\mathbf{N}} )^2  - \frac{4}{\theta_{\mathbf{\delta}}}
  N_{\mathbf{\delta}}
  [  1 - \cos (u_{\mathbf{N}}  -
  u_{\mathbf{N} - \mathbf{\delta}} ) ]\right\} \; ,
\end{equation}
where $\mathbf{N}$ is a vector formed of $k$ oscillator level number operators,
taking values in $k$-dimensional lattice, $\mathbf{\delta}$ is a unite lattice
vector, $N_{\mathbf{\delta}}$ is the $\mathbf{\delta}$-th component of $\mathbf{N}$
and $\theta_{\mathbf{\delta}}$ is the $\mathbf{\delta}$-th diagonal value of
noncommutativity matrix (we assume that $\theta^{\mu\nu}$, is brought to the
block-diagonal form), $\theta^k\equiv \theta_1\dots\theta_k$. Up to the factor
$N_{\mathbf{\delta}}$ this system coincides with the system of coupled rotators on
the $k$-dimensional lattice \cite{Polyak}, but living on a quadrant $N_\delta\geq
0$.

The above equations are operator ones. By passing to the oscillator basis
$\{\ket{n}\}$: $N\ket{n}=n\ket{n}$ the action (\ref{action_ansatz}) and equation
of motion (\ref{ansatz}), are rewritten in a simple lattice form,
\begin{align}\label{ans_lat}
  &\ddot{u}_n =  \frac{2}{\theta} \left[(n+1)
  \sin (u_{n+1} - u_n  ) - n \sin (u_n  - u_{n-1} ) \right] \quad (n \in \mathbb{Z}_+), \\ \label{action_lat}
  &S = \frac{2\pi\theta}{\lambda^2}  \int_{-\infty}^{+\infty} \dd t  \sum_{n=0}^\infty
  \left\{\frac{1}{2}(\dot{u}_n )^2 - \frac{4}{\theta} n  [  1 - \cos (u_n  -
  u_{n-1} ) ]\right\},
\end{align}
suitable for the analysis. If, e.g. one sets naively $u_n=0$, for $n>0$, one gets
the sin-Gordon equation for $u_0$.

Let us remark that the system (\ref{ans_lat},\ref{action_lat}) has an interesting
algebraic meaning. In order to see it let us rewrite eq. (\ref{ans_lat}) in the
form as follows,
\begin{gather}
 \ddot{u}_0 = \frac{2}{\theta}  \sin (u_0  - u_{1} ) ,
\qquad \pd^2_t (u_1 + u_0) = \frac{2}{\theta} 2 \sin (u_1  - u_2 ),\qquad \dots \\
\label{8} \pd^2_t \left( \sum_{n=0}^{k} u_n \right) = \frac{2}{\theta} \, (k+1) \,
\sin (u_k - u_{k+1} ),
\end{gather}
obtained by summing up the equations with labels from zero to some $k$.

The conservation of the full momentum requires $\pd^2_t ( \sum_{n=0}^{\infty} u_n )
= 0$. Due to summation in eq. (\ref{8}), the momentum is convergent only if
$(u_{k+1} - u_k)$ decreases faster than $1/k$.

It is possible to rewrite the equations (\ref{8}) in the form as follows,
\begin{equation}\label{9}
  \pd^2_t ( v_k ) = \frac{2}{\theta} \, (k+1) \, \sin \left( \sum_l \, a_{kl} v_l \right),
\end{equation}
where $v_k = \sum_{n=0}^{k} u_n$ and $a_{kl}$ is a Cartan matrix for the
sl$(\infty)$ algebra
\begin{equation}\label{10} a_{kl} = \left(
  \begin{array}{cccc}
    2 & -1 & 0 & \dots \\
    -1 & 2 & -1 & \dots \\
    0 & -1 & 2 & \dots \\
    . & . & . & ...
  \end{array}
\right)
\end{equation}

Let us return back to the model (\ref{ans_lat}), (\ref{action_lat}). As one can
see, the equations of motion possess following static solutions,
\begin{equation}\label{n_vac}
  u_n^{\mathrm{vac}}=\pi m_n,
\end{equation}
where $m_n$ are integers. Solutions with all $m_n$ even give true (stable) vacua
of the model, while the presence of odd $m_n$ corresponds to excitations over
these vacua.

Let us note, that the r.h.s of the eq.(\ref{ans_lat}) is lattice analog of an
elliptic operator, i.e. any static solution to the eq (\ref{ans_lat}) reduces to
(\ref{n_vac}).\footnote{However, one can find a nontrivial uniformly accelerating
solution, $u_n(t)=u^{(0)}_n+(a/2)t^2$, where $\dot{u}^{(0)}_n=0$.}

However, in the case of ``spherical symmetry'' one is justified to consider
solutions with punched point at the origin. This eliminates the first equation
from the system (\ref{ans_lat}). The remaining equations look as follows,
\begin{equation}\label{psd_vac_eq}
  \frac{2}{\theta} \left[ n \sin (u_n  - u_{n-1} ) - (n+1)
  \sin (u_{n+1} - u_n  ) \right]=0,\qquad n=1,2,\dots
\end{equation}
The general solution to eq. (\ref{psd_vac_eq}) is given by,
\begin{equation}\label{psd_vac_sol}
  u_n=u_0 +\sum_{k=1}^n
  (-1)^{m_k-m_{k-1}}\arcsin
  \left(\frac{s_1}{k}\right)+\pi m_n,
\end{equation}
where $u_0$, and $s_1$ are constants of ``integration'', $-\pi\leq u_0\leq\pi$,
$-1\leq s_1\leq 1$. Again, in the case when all $m_n$ are even this solution
provides a local minimum for the energy, i.e. a metastable state. Conversely if
there are odd $m_n$ this is an unstable state.

The solution of the type (\ref{psd_vac_sol}), is the lattice analog of Coulomb
potential. One can readily generalise the above solution to the case of arbitrary
dimension $p$ of the noncommutative space. Indeed, the ``spherical symmetry'' means
that solution should depend only on the sum of all $N_\delta$,
$\delta=1,\dots,p/2$ and it is given by,
\begin{equation}\label{d_dim}
  u_n=u_0 +\sum_{k=1}^n (-1)^{m_k-m_{k-1}}
  \arcsin
  \left(\frac{(k-1)!(D-1)!}{(k+D-1)!}s_1\right)+\pi m_n,
\end{equation}
once again, $u_0$ and $s_1$ are constants of integration.

It is worthwhile to note that due to the long range character of the Coulomb force
the energies corresponding to the solutions (\ref{psd_vac_sol}), (\ref{d_dim})
diverge.

When $\theta\to 0$, the above solutions approach the Coulomb potential respectively
in two and $p$ dimensions. However, there are some subtleties connected with this
limit which we consider in Section 5.
\section{$U$-field on a special $q$-space}

As we have seen above, the equations we met, include sums over an infinite tower of
$n$, which potentially may diverge. Also, if one wants to apply numerical analysis
to our system one face the problem that the brute truncation of the tower by some
finite $n_{\mathrm{f}}$ is inconsistent. However, one can generalise the above
construction by letting fields live on a special $q$-space instead of
noncommutative plane. This $q$-space (when $q$ is $2n_{\mathrm{f}}$-th root of
unity) can be used for a regularisation of the model, because the corresponding
algebra has finite dimensional representations and approaches the algebra of the
noncommutative plane for $n_{\mathrm{f}}\to\infty$.

The mentioned $q$-space is constructed as follows. Consider more general algebra
of quantum oscillator \cite{q-alg}, \cite{Filippov:1991my} (which is related to
the $q$-deformed Heisenberg algebra \cite{Wess}), generated by $A$ and $\bar{A}$,
\begin{equation}\label{osc1}
  A \, \bar{A} - q \bar{A} \, A = q^{-N}, \qquad A = a \, \sqrt{\frac{[N]_q}{N}},
  \quad \bar{A} = \sqrt{\frac{[N]_q}{N}}  \bar{a} ,
\end{equation}
where $[N]_q = (q^N - q^{-N})/(q - q^{-1})$, while $a$ and $\bar{a}$ correspond to
usual oscillator. For $q \to 1$ we get $A \to a$ and $\bar{A} \to \bar{a}$.

One can, therefore, find
$$
  \bar{A} A = [N]_q , \qquad A \bar{A} = [N+1]_q \; .
$$
Define the derivatives with respect to noncommutative coordinates, as
follows
$$
  \pd_2 (\cdot) =- \ii \sqrt{\frac{1}{2 \theta}}[( A+\bar{A} ), (\cdot) ],
  \quad \pd_1 (\cdot) = \sqrt{\frac{1}{2 \theta}}
  [( A -  \bar{A}) , (\cdot) ].
$$

For $q$ being root of unity ($q^{2n_{\mathrm{f}}} =1$) one can put
$A^{n_{\mathrm{f}}} = 0 = (\bar{A})^{n_{\mathrm{f}}}$. The action for the quantum
oscillator system ($q$ is arbitrary) is chosen in the form
\begin{multline}\label{osc2}
  S = -\frac{2\pi\theta}{\lambda^2}  \int_{-\infty}^{+\infty} \dd t\times
  \\ \sum_{n=0}^{\infty}
  \bra{n} \, \left\{ \pd_t U_N \pd_t U^{-1}_N  - \frac{1}{\theta}
  ([\bar{A} , U_N ]  [ A, U^{-1}_{N}] + [ A,U_{N} ] [
  \bar{A},U^{-1}_N ]) \right\} \ket{n} = \\
  = -\frac{2\pi\theta}{\lambda^2} \int_{-\infty}^{+\infty} \dd t
  \sum_{n=0}^{\infty}
  \left\{ \pd_t U_n \pd_t U^{-1}_n  - \frac{2}{\theta} [n+1]_q
  (U^{-1}_n  U_{n+1} + U^{-1}_{n+1} U_n - 2 ) \right\}.
\end{multline}

The equations of motions corresponding to this model can be written in the form
analogous to (\ref{8})
\begin{gather}\label{osc3}
  \pd^2_t u_0 = \frac{2}{\theta}  \sin (u_0  - u_{1} ), \quad
  \pd^2_t (u_1 + u_0) = \frac{2}{\theta} 2 \sin (u_1  - u_2 ), \quad \dots\\
  \pd^2_t \left(\sum_{n=0}^{k} u_n \right) = \frac{2}{\theta} \, [k+1]_q \sin (u_k  -
  u_{k+1} ).
\end{gather}

or, equivalently,
\begin{equation}\label{osc4}
  \pd^2_t ( v_k ) = \frac{4}{\theta} \, [k+1]_q
  \sin (a_{kl} v_l ),
\end{equation}
where $v_k = \sum_{n=0}^{k} u_n$ and $a_{kl}$ is the Cartan matrix (\ref{10}).
Since $[n_{\mathrm{f}}]_q = 0$ for $q^{2n_{\mathrm{f}}} =1$, the system of
equations (\ref{osc3}) or (\ref{osc4}) will be cut for $k = n_{\mathrm{f}}-1$. In
this case we have the finite chain of interaction point particles, suitable e.g.
for numerical analysis.

The analog of solution (\ref{psd_vac_sol}) in the $q$-deformed space looks as
follows,
\begin{equation}\label{q_psd}
  u_n=u_0 +\sum_{k=1}^n (-1)^{m_k-m_{k-1}}\arcsin \left(\frac{s_1}{[k]_q}\right)
  +\pi m_{n},
\end{equation}
with some constants $u_0$ and $s_1$. Due to finite size of the
lattice in this case this solution is a finite energy one.
\section{Continuum limit and Polyakov vortices}

In this section we consider $p=4$ with nondegenerate noncommutativity matrix
$\theta_{\mu\nu}$, $\mu,\nu=1,\dots,4$, and no time, i.e. we assume the Euclidean
signature. Without loss of generality consider $\theta_{\mu\nu}$ is block diagonal.
The commutators of $x^\mu$ are given by,
\begin{align}\label{4d_comm}
  &[x^1,x^2]=\ii \theta_{(1)} \\
  &[x^3,x^4]=\ii \theta_{(2)}, \qquad \theta_{(\delta)}>0.
\end{align}

Let us study the model describing excitations radial in $(x^1,x^2)$ and $(x^3,x^4)$
planes. The action is given by energy corresponding to eq. (\ref{action2k}) for
$(p/2)=2$,
\begin{equation}\label{action_4d}
  S=(2\pi\theta)^2\sum_{\mathbf{n}}
  \frac{4}{\lambda^2\theta_{\mathbf{\delta}}}n_{\mathbf{\delta}}(1-\cos(u_{\mathbf{n}}-
  u_{\mathbf{n}-\mathbf{\delta}})).
\end{equation}
In the absence of time coordinate this is a purely lattice action with dimensional
lattice spacing parameter equal to $(2\pi \theta)$. The factor $(2\pi\theta)^2$ in
the front of the sum is the volume of the elementary cube.

Consider the partition function describing such excitations,
\begin{equation}\label{z}
  Z=\int_{-\pi}^{\pi}\prod_{\mathbf{n}}\frac{\dd
  u_{\mathbf{n}}}{2\pi}\e^{-S}.
\end{equation}
As it was already mentioned the action (\ref{action_4d}) (modulo the factor
$(1/\theta_{\mathbf{\delta}}) n_{\mathbf{\delta}}$), describes interaction of
rotators on a two-dimensional lattice \cite{Polyak}. The presence of factor
$n_\delta$, can be interpreted as the nontrivial metric background on the lattice.
The exact meaning of this factor will become clear when we will take the continuum
(i.e. commutative) limit.

In fact the analysis of Ref. \cite{Polyak}, applies also here. Indeed, consider the
isotropic weak coupling limit $\beta\equiv 4/(\lambda^2\theta)\to\infty$, of the
model (\ref{z}); we set $\theta_{(1)}=\theta_{(2)}\equiv\theta$. In this limit the
main contributions to the partition function come from configurations where
neighbour $u_{\mathbf{n}}$ are close modulo $2\pi\times$(integer). Following
Polyakov let us introduce field $m_{\mathbf{n},\mathbf{\delta}}$ to describe this
integer factor and use the trick to substitute the partition function (\ref{z}) by
\begin{equation}\label{Z_m}
  Z=\sum_{\{m_{\mathbf{n},\mathbf{\delta}}\}}\int_{-\pi}^{\pi}\prod_{\mathbf{n}}
  \frac{\dd u_{\mathbf{n}}}{2\pi}\exp\left(-\frac{\beta}{2}(2\pi\theta)^2
  \sum_{\mathbf{n},\mathbf{\delta}}
  n_{\mathbf{\delta}}(u_{\mathbf{n}}-u_{\mathbf{n}-\mathbf{\delta}}
  +2\pi m_{\mathbf{n},\mathbf{\delta}})^2\right).
\end{equation}

Using the lattice analog of the Hodge decomposition theorem
\cite{Berezinsky:1971fr,Kosterlitz:1973xp} (see also \cite{Seiler}), one can
decompose the integer vector field $m_{\mathbf{n},\mathbf{\delta}}$ into gradient
and vortical parts as follows,
\begin{equation}\label{Hodge}
  m_{\mathbf{n},\mathbf{\delta}}=\alpha_{\mathbf{n}}-\alpha_{\mathbf{n}-\mathbf{\delta}}
  +\tilde{n}_{\tilde{\mathbf{\delta}}}(\eta_{\tilde{\mathbf{n}}+\tilde{\mathbf{\delta}}}-\eta_{\tilde{\mathbf{n}}}),
\end{equation}
where the tilde refers to the dual lattice, and the factor
$\tilde{\mathbf{n}}_{\tilde{\mathbf{\delta}}}$ in the last term comes from the
nontrivial metric. The part containing $\alpha_{\mathbf{n}}$ carries the gradient
part while the factor $\eta_{\tilde{\mathbf{n}}}$ the vortical one,
\begin{multline}\label{vort}
  \sum_{\mathrm{plaquette}}m_{\mathbf{n},\mathbf{\delta}}\equiv
  m_{n,\delta_1}+m_{n+\delta_1,\delta_2}-
  m_{n+\delta_2,\delta_1}-m_{n,\delta_2}\\
  =\sum_{\delta}((2\tilde{n}_\delta+1)\eta_{\tilde{\mathbf{n}}}
  -\tilde{n}_\delta\eta_{\tilde{\mathbf{n}}-\delta}-
  (\tilde{n}_\delta+1)\eta_{\tilde{\mathbf{n}}+\delta})
  \equiv q_{\tilde{\mathbf{n}}}.
\end{multline}
The first line is the lattice analog of external derivative,  while the second
line is the lattice analog of the Laplace operator. The vorticity source living on
the dual lattice is denoted as $q_{\tilde{\mathbf{n}}}$.

Making the field redefinitions one can absorb the gradient part
$\alpha_{\mathbf{n}}$ into the field $u_{\mathbf{n}}$, spreading its values from
the interval $(-\pi,\pi]$ to $(-\infty,+\infty)$. The partition function becomes,
\begin{multline}\label{4.30}
  Z=\sum_{\{q_{\tilde{\mathbf{n}}}\}}\int_{-\infty}^{+\infty}\prod_{\mathbf{n}}
  \frac{\dd u_{\mathbf{n}}}{2\pi} \\
  \times\exp\left(-\frac{\beta}{2}(2\pi\theta)^2
  \left(\sum_{\mathbf{n},\mathbf{\delta}}
  n_\delta(u_{\mathbf{n}}-u_{\mathbf{n}-\mathbf{\delta}})^2+
  2\pi^2\sum_{\tilde{\mathbf{n}},\tilde{\mathbf{\delta}}}n_{\delta}
  (\eta_{\tilde{\mathbf{n}}}-\eta_{\tilde{\mathbf{n}}-
  \tilde{\mathbf{\delta}}})^2\right)\right)\\
  =Z_{\mathrm{Gauss}}\sum_{\{q_{\tilde{\mathbf{n}}}\}}
  \exp\left(-\frac{\beta}{2}(2\pi\theta)^2\sum_{\tilde{\mathbf{n}},\tilde{\mathbf{n}}'}
  q_{\tilde{\mathbf{n}}}\Delta^{-1}_{\tilde{\mathbf{n}},\tilde{\mathbf{n}}'}
  q_{\tilde{\mathbf{n}}'}\right)
\end{multline}
where,
\begin{equation}\label{z_gauss}
  Z_{\mathrm{Gauss}}=\prod_{\mathbf{n}}\int_{-\infty}^{+\infty}
  \frac{\dd u_{\mathbf{n}}}{2\pi}\exp \left(-\frac{\beta}{2}(2\pi\theta)^2
  \sum_{\mathbf{n},\mathbf{\delta}}
  n_\delta(u_{\mathbf{n}}-u_{\mathbf{n}-\mathbf{\delta}})^2\right)
\end{equation}
and $\Delta^{-1}$ is the inverse operator to the Laplace operator $\Delta$ given
by,
\begin{equation}\label{Delta}
  (\Delta \eta)_\mathbf{n}\equiv\sum_{\delta}\left(
  (2n_\delta+1)u_{\mathbf{n}}-nu_{\mathbf{n}-\delta}
  -(n_\delta+1)u_{\mathbf{n}+\delta}\right).
\end{equation}

The part described by $Z_{\mathrm{Gauss}}$ gives the ``naive'' weak coupling limit
while the remaining part describes the gas of Polyakov vortices. Configurations
with such vortices are suppressed in the limit $\beta\to\infty$. Recall that
$\beta\sim 1/\theta$, therefore this is both continuum and commutative limit.

In this limit one can pass from discrete notation to continuous ones $u_n\to
u(x)$, where $x^\delta=(2\pi\theta) n^\delta$. Indeed, the fact that the quantity
$u_n-u_{n-\delta}$ is small allows to do this. In these notations the action for
the field $u$ becomes,
\begin{equation}\label{cont_action}
  S_{\mathrm{cont}}= \frac{1}{\lambda^2_{\mathrm{cont}}}\int\dd^2 x \left(
  x^1(\pd_1u)^2+x^2(\pd_2u)^2\right),
\end{equation}
which after the change of coordinates,\footnote{Remember that $x^\delta\geq 0$.}
$x^\delta\to r^\delta=\sqrt{x^\delta}/(2\pi)$, looks like four-dimensional action
for the excitations which are radial in the planes (1,2) and (3,4),
\begin{equation}\label{radial}
  S_{\mathrm{cont}}=\frac{1}{\lambda^2_{\mathrm{cont}}}\int
  2\pi \dd r^1 r^1\,2\pi \dd r^2 r^2 \left(
  (\pd_{r^1}u)^2+(\pd_{r^2}u)^2\right),
\end{equation}
where $r^1$, and $r^2$ play the r\^{o}le of radii in the planes
(1,2) and (3,4) respectively.

In the weak coupling limit the the configurations with vortices enters with a
factor $\exp(-\beta \times\mathrm{constant})$, and are exponentially suppressed.
In the strong coupling limit the r\^ ole of vortices become important.
\section{Discussions}

In this paper we considered noncommutative charged tachyonic field. We have shown,
that the vacuum dynamics of this field is described by the noncommutative
$U$-field.

The model under consideration exhibits a tight relation to conventional lattice
models. The similar relation was met also in the IKKT model, which is closely
related to the noncommutative (super)Yang--Mills model \cite{Sochichiu:2000fs},
where it was shown the fermionic doubling, a phenomenon well known on the lattice.

So far, we considered both classical solutions and partition function for
excitations which possess the noncommutative analog of spherical symmetry.

We have shown, that the system of these excitations is equivalent to one of
rotators on an inhomogeneous lattice. This system has a reach structure of vacua.
The limit of small noncommutativity corresponds to the weak coupling limit for the
rotators. Thus, in the case of four Euclidean dimensions, this limit yields  a free
scalar field surrounded by the gas of exponentially suppressed vortices. As the
noncommutativity parameter increases the vortex gas become more important, which
may signal phase transitions. Since this corresponds to the string limit, perhaps
the vortices can be identified with fundamental strings.

It is interesting to note that, with departing the origin the effective
interaction weakens regardless how strong it was at origin. Therefore, far enough
from origin one \emph{always} has weak coupled regime, which means that the
vortices cannot depart far from there.

We analysed the radial part of the noncommutative sigma model. It would be
interesting to extend this analysis also to the noncommutative analogs of spherical
harmonics (of course, if one succeed to define them).

Also, in the light of a recent paper \cite{Gerasimov:2000ga}, the Higgs mechanism
may play some important r\^ ole in the vacuum structure of the string field theory,
which in the low energy limit reduces to the noncommutative Higgs model considered
in Ref. \cite{Jatkar:2000ei,Harvey:2000jb}, to which our model is relevant.
\acknowledgments We are grateful to A.~Pashnev, S.~Sergeev and K.~Zarembo for
helpful discussions. Work of A.P.I. was supported by RFBR grant \# 00-01-00299,
work of C.S. by RFBR grant \# 99-01-00190, INTAS grant \# 950681, and Scientific
School support grant 96-15-0628, and M.T. was supported in part by the RFBR grant
\# 99-02-18417.
\providecommand{\href}[2]{#2}\begingroup\raggedright\endgroup
\end{document}